# Temperature response of the neuronal cytoskeleton mapped via atomic force and fluorescence microscopy

*Elise Spedden[1], David L. Kaplan [2], and Cristian Staii[1,*]*

[1]Department of Physics and Astronomy and Center for Nanoscopic Physics, Tufts University, 4 Colby Street, Medford, MA 02155

[2]Department of Biomedical Engineering, and Department of Chemical Engineering, Tufts University, 4 Colby Street, Medford, MA, 02155

[*] CORRESPONDING AUTHOR: Cristian Staii, Department of Physics and Astronomy, Tufts University, 4 Colby Street, Medford, MA 02155, USA Ph.: (617) 627-5368.

E-mail: cristian.staii@tufts.edu

**Abstract**

Neuronal cells change their growth properties in response to external physical stimuli such as variations in external temperature, stiffness of the growth substrate, or topographical guidance cues. Detailed knowledge of the mechanisms that control these biomechanical responses is necessary for understanding the basic principles that underlie neuronal growth and regeneration. Here, we present elasticity maps of living cortical neurons (embryonic rat) as a function of temperature, and correlate these maps to the locations of internal structural components of the cytoskeleton. Neurons display a significant increase in the average elastic modulus upon a decrease in ambient temperature from 37°C to 25°C. We demonstrate that the dominant mechanism by which the elasticity of the neurons changes in response to temperature is the stiffening of the actin components of the cytoskeleton induced by myosin II. We also report a reversible shift in the location and composition of the high-stiffness areas of the neuron cytoskeleton with temperature. At 37°C the areas of the cell displaying high elastic modulus overlap with the tubulin-dense regions, while at 25°C these high-stiffness areas correspond to the actin-dense regions of the cytoskeleton. These results demonstrate the importance of considering temperature effects when investigating cytoskeletal dynamics in cells.

1. Introduction

A detailed knowledge of the processes by which neuronal cells change their elastic properties in response to external stimuli is essential for understanding neuronal growth and the development of the nervous system. The elastic modulus and mechanical integrity of neurons are largely dependent on the cytoskeleton, a deformable and dynamic polymer network including actin filaments, intermediate filaments, and microtubules, which provides structure, support and signaling within the cell body and the cellular extensions (neurites). The cytoskeletal components exhibit rearrangements in response to many types of stimuli such as axonal outgrowth, variations in the local extracellular environment, variations in temperature, generation of traction forces, and mechanical interactions between neurons or between neurons and their growth substrates [1-6]. Measuring the elastic modulus of neurons under these dynamic conditions could therefore lead to a deeper understanding of the mechanisms that control neuronal architecture and structural integrity during growth and development, or during nerve recovery after injury or traumatic damage.

Atomic Force Microscopy (AFM) is a very powerful technique for measuring mechanical properties of living cells. The AFM can be used to acquire high-resolution images of the cell topography under physiological conditions, to apply forces with well-controlled magnitude and orientation, to map the elastic modulus of living cells with high spatial resolution and to track changes in elastic modulus over time [6-12]. When combined with an optical stage and a temperature-controlled environment, the AFM can be used to map the elastic modulus of single living cells over time and under varying temperatures and external conditions [13], while the cell morphology is simultaneously monitored via optical microscopy, and the intracellular components are tracked via fluorescence microscopy [6]. Knowing which cytoskeletal components dominate the elastic landscape of a cell can provide a better understanding of mechanical interactions and cellular response to the surrounding environment, as well as better predictions for drug responses.

Understanding how the cellular elastic properties depend on temperature can yield new insight into the mechanisms of structural support and mechanical response within a given cell type [13]. Furthermore, since the mechanical properties of cells vary with temperature [13-16] it is very important to consider the effects of temperature on the results when comparing measured elastic moduli between different cell types, or among cells of a given type measured at different temperatures. For example, while some recent measurements of the cellular elastic modulus have been performed at or around 37°C to approximate the physiological environment of the cell types measured (rat cortical neurons) [6, 7, 17], many previous measurements on various other cell types were performed at ambient temperature (at or around 25°C) [11, 12, 18, 19]. Finally, a detailed mapping of temperature-dependent effects on cellular structure and mechanics can also provide insight into physiologically relevant changes. For example, local tissues undergo a drop in physiological temperature during surgery, while an increase in temperature occurs during infection or a fever. Such localized changes in temperature *in vivo* may have consequences for cell structure and thus function, yet these temperature effects are currently poorly understood.

Temperature-dependent elastic modulus measurements have been previously made on several cell types including alveolar and breast epithelial cells, fibroblasts, and human bone marrow-derived mesenchymal stem cells (hMSCs). Alveolar epithelial cells showed a distinct increase in elastic modulus when cells were warmed from room to physiological temperature [13], whereas fibroblast [14], breast epithelial [15], and hMSCs [16], as well as rat cortical neurons in this study, showed a decrease in the cell's elastic modulus with increasing temperature. To date the causes of these variations in cellular elasticity with temperature are not yet elucidated [14-16].

Cortical neurons display low values for the measured elastic modulus when compared to most cell types in the body, and they also grow in one of the softest environments *in vivo* [4, 6, 18, 20]. In addition, the neuron elastic properties are very sensitive to external physical stimuli (stiffness of the environment, topography and temperature) [1, 2, 4, 8] and thus it is very important to understand the dynamics of the cytoskeletal components in response to these physical stimuli. To date there are no studies that correlate changes in the elastic properties of neurons with specific changes in the cytoskeletal components due to variations in temperature.

In this paper, we determine how the elastic modulus of cortical neurons varies with temperature. We use AFM imaging and AFM-based force mapping to obtain maps of the cell topography and elastic modulus and to measure how these maps change with respect to temperature. Our results show an increase in the elastic modulus of cortical neurons with a drop in ambient temperature from 37°C to 25°C. In addition, we use combined AFM – based elasticity and fluorescence microscopy measurements to correlate the elastic modulus maps to the locations of internal components of the cytoskeleton. We report a switch in the elastically dominant cytoskeletal features from tubulin-dense regions at 37°C to actin-dense regions at 25°C. Furthermore, through targeted chemical modification of the cells we demonstrate that the dominant mechanism by which the cell stiffness is changing is through a temperature-dependent change in myosin II dynamics in cortical neurons.. This temperature-induced shift in cytoskeletal mechanical properties has implications in terms of neuron functions *in vivo* in response to infections or tissue regeneration, as well as for the ability to manipulate cell outcomes *in vitro*.

**2. Materials and Methods**

*2.1 Surface preparation, cell culture and plating*

Rat cortices were obtained from embryonic day 18 rats (Tufts Medical School). The brain tissue isolation protocol was approved by Tufts University Institutional Animal Care and Use Committee and complies with the NIH Guide for the Care and Use of Laboratory Animals. The cortical tissue was incubated at 37ºC in 5 mL of trypsin for 20 minutes. Trypsin was inhibited with 10 mL of neurobasal medium (Life Technologies, Frederick, MD) which was supplemented with GlutaMAX, b27 (Life Technologies), pen/strep, and 10 mg of soybean trypsin inhibitor (Life Technologies). Mechanical dissociation of the rat cortices was performed, the cells were centrifuged, the supernatant was removed, and the cells were resuspended in 20 mL of neurobasal medium with L-glutamate (Sigma-Aldrich, St. Louis, MO). The neurobasal media was used to support neuronal growth without serum, a technique which reduces glial cell proliferation [6]. The cells were mechanically re-dispersed, counted, and plated at a density of 250,000 cells per 3.5 cm culture disk. Each sample of cells was grown for a minimum of 2 days before measurements. Immunostaining experiments performed on this type of samples indicate cultures of high neuron purity [6]. Cell samples were cultured on 3.5 cm glass disks manufactured to fit in the Asylum Research Bioheater fluid cell (Asylum Research, Santa Barbara, CA). Poly-D-lysine (PDL) (Sigma-Aldrich, St. Louis, MO) coating was added to the glass disks by immersing them in a PDL solution (0.1 mg/ml) for 2 hours at room temperature. The disks were rinsed twice with sterile water, and sterilized using ultraviolet light for at least 30 minutes.

*2.2 Force map acquisition and data analysis*

Force Maps were performed on an Asylum Research MFP-3D-Bio AFM (Asylum Research, Santa Barbara, CA) integrated with an inverted Nikon Eclipse Ti optical microscope (Micro Video Instruments, Avon, MA). Each sample was mounted in an Asylum Research Bioheater chamber with 1 ml of cell culture medium. The samples were maintained at either 25°C or 37°C during each experiment. All experiments were carried out in the temperature controlled chamber, which was sealed against evaporation. The $CO_2$ concentration was not controlled during the experiments. During each temperature change on a single sample, the new measurements were not taken until the sample had reached and consistently maintained the new temperature state for a minimum of 10 minutes. All measurements were performed with Olympus Biolever cantilevers (Asylum Research, Santa Barbara, CA) possessing a nominal spring constant of 0.03 N/m. To verify that all the cells to be measured were alive, time lapse videos (typically of 10 minutes) were taken of each cell set before AFM measurements. The average lifetime of a neuronal cell in the AFM Bioheater fluid chamber is approximately 4 hrs, and all AFM and fluorescent measurements were taken within the first two hours of the sample being introduced into the chamber. All measured cells had similar soma size, with an average diameter of (11 ± 3) µm. All force measurements were performed only on the neuron soma. Before measurements were performed on each new sample, the cantilever was calibrated first in air and then in the sample medium. 16 X 16 µm maps of individual force vs. indentation curves were taken on each cell with a resolution of 1 µm between points as described previously [6]. To limit energy dissipation due to viscoelastic effects the indentation frequency was 0.33 Hz. All indentation settings were kept consistent for all measurements: maximum force was set to 0.2 (+/- 0.05) nN which resulted in a maximum indentation between 1 and 2 µm. Each map took approximately 15 minutes to complete. The elastic modulus values were determined by fitting the Hertz model for a 30 degree conical indenter to the acquired force vs. indentation curves using the Asylum Research MFP-3D Hertz analysis tools, as previously reported [6]. A typical force point has a fitting error of ≤ 20% and the values obtained for the elastic modulus at every point are reproducible within this error. Force curves that were obtained over the substrate rather than over the cell body were excluded from processing using topographical information obtained during mapping, and the remaining data was analyzed for each map using the Asylum Research MFP-3D software. A table comparing the absolute mean values of the elastic moduli for the cells at different conditions is given in the supplementary materials (table T1).

*2.3 Fluorescence microscopy and chemical modification of the neurons*

All fluorescent measurements were performed on the inverted Nikon Eclipse Ti optical stage, integrated with the Asylum Research AFM using either the 20X or 40X objectives. We utilized two different fluorescent stains in this study. For microtubule staining, live cortical neurons were rinsed with phosphate buffered saline (PBS) and then incubated at 37ºC with 50 nM Tubulin Tracker Green (Oregon Green 488 Taxol, bis-Acetate, Life Technologies, Grand Island, NY) in PBS. After staining, each sample was rinsed twice then transferred to the AFM Bioheater chamber where it was maintained at fixed temperatures (37ºC or 25ºC) for the acquisition of bright field and fluorescence images (using a standard Fluorescein isothiocyanate -FITC filter: excitation/emission of 495 nm/521 nm).

For F-actin staining we have located optically each set of live cortical cells on the surface. We acquired force maps for two to three cells per cultured substrate at fixed temperatures (37ºC or 25ºC). Immediately after the last map was taken, the sample was removed from the AFM stage and fixed in 10% formalin, and maintained at either 25ºC or 37ºC for 15 minutes. The sample was then rinsed with PBS and permeabilized with 0.1% Triton-100 (Sigma-Aldrich, St. Louis, MO) in phosphate-buffered saline (PBS) (Life Technologies, Grand Island, NY) for 10 minutes, then incubated at room temperature for 20 minutes in 150nM Alexa Fluor® 564 Phalloidin (Life Technologies, Grand Island, NY), and finally rinsed with PBS. The cells originally mapped via AFM were relocated optically using sample markings and imaged fluorescently using a standard Texas Red (excitation/emission of 596/615) filter.

Taxol treated cells were incubated for 24 hours with 10 μM of paclitaxel (generic name for Taxol, Life Technologies, Grand Island, NY) in cell media then transferred to the AFM bioheater for temperature-dependent force mapping. Taxol stabilizes microtubule dynamics [21] and suppresses the temperature-dependence of microtubule flexural rigidity [22]. Blebbistatin treated cells were initially plated with 10 μM Blebbistatin (Sigma-Aldrich, St. Louis, MO) in the cell media and incubated for a minimum of 2 days before measurement. Blebbistatin disrupts actin filaments through myosin II inhibition, increasing instability in contractile actin-myosin interactions within the cell [23, 24]. Y-27632 treated cells were incubated for 1 hour with 10 μM of Y-27632 (Sigma-Aldrich, St. Louis, MO) in cell media before the measurements. Adding Y-27632 has been shown to disrupt actin bundles and inhibit reorganization of actin-based structures [25]. Myosin II-based formation of contractile F-actin structures such as stress fibers is also inhibited by the addition of Y-27632 [13, 26, 27].

## 3. Results

*3.1 Variation of the neuron elastic modulus with temperature*

Force maps were acquired on individual, chemically unmodified cortical neuron soma at both 37ºC and 25ºC. The cells were monitored optically during all force map acquisitions to rule out any cells with visible indicators of unhealthy change, such as spreading or collapse of the soma, rapid neurite retraction, or blebbing (see section 2.2). It is also important to emphasize that according to our previous results the active growth of neurites at 37ºC generates a substantial increase in cell elasticity due to tubulin aggregation [6]. These phases of active neurite growth lead to a substantial shift in the measured cell elastic modulus towards higher values, and therefore could potentially interfere with the temperature-induced variations in cell stiffness (see below). To minimize this effect and isolate the temperature-dependent response, in this paper we focus exclusively on the results obtained for the cells that *have not* displayed any active growth of neurites during the elasticity measurements for all temperatures (N=25 cells).

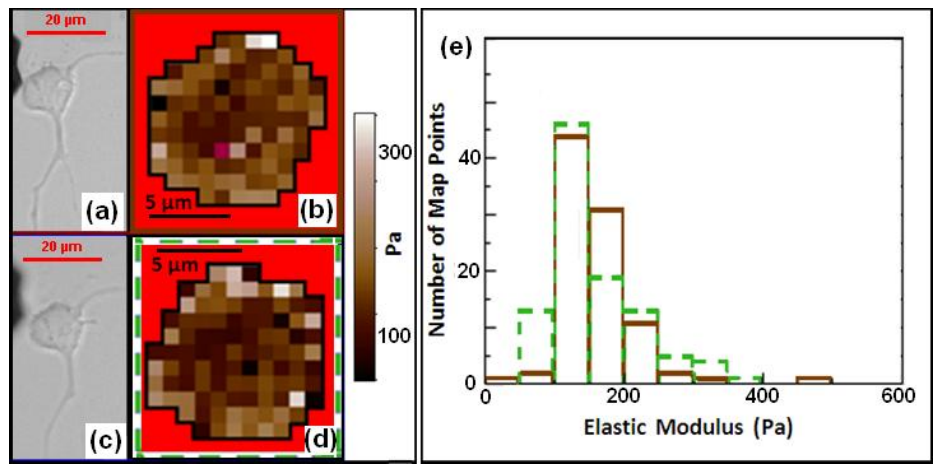

**Figure 1**. Reproducibility of the elastic modulus maps over time. (a) Optical image of a chemically unmodified live cortical cell, which is not undergoing active neurite growth. (b) Elasticity map for the cell shown in (a) measured at 37ºC. (c) Optical image of the same cell as in (a) and taken 45 minutes after acquiring the elasticity map shown in (b). (d) Subsequent elasticity map of the same cell as shown in (b), measured after 45 minutes and in the same conditions (37ºC, no active neurite extension, no chemical modifications). (e) Histograms of elastic modulus for the elasticity maps shown in (b) (solid brown line) and (d) (dotted green line). The average elastic modulus values between the two maps differ by less than 2% (from 160 Pa to 158 Pa). Similar data was obtained

for 5 other unmodified cells measured at 37ºC, and 7 other cells measured at 25ºC (see supplementary figure S1 for an example of a cell at 25ºC). All elasticity maps were measured between 20 minutes- 2 hrs apart. The % difference between the average elastic modulus for subsequent maps measured for the same cell and at the same temperature varied between 1% and 10%.

Figure 1 demonstrates the reproducibility of the force maps at a given temperature. Figures 1(b) and 1(d) show two examples of elasticity maps for the same cell measured at 37ºC and taken 45 minutes apart (the corresponding optical images are shown in figures 1(a) and 1(c), respectively). The histograms of measured elastic moduli are given in figure 1(e) (brown continuous line corresponds to map shown in figure 1(b), and dotted green line corresponds to map shown in figure 1(d), respectively). The data shows that although there are some small changes in the distribution of the map points, the average elastic modulus between the two maps differs by less than 2%. We find that repeated measurements of elastic modulus for single healthy cells (N=13 cells total) yield overlapping elasticity maps, with variations in the average elastic modulus for a single cell of less than 10%, over the course of 20 minutes - 2 hrs, thus demonstrating the reproducibility of the force maps at a given temperature.

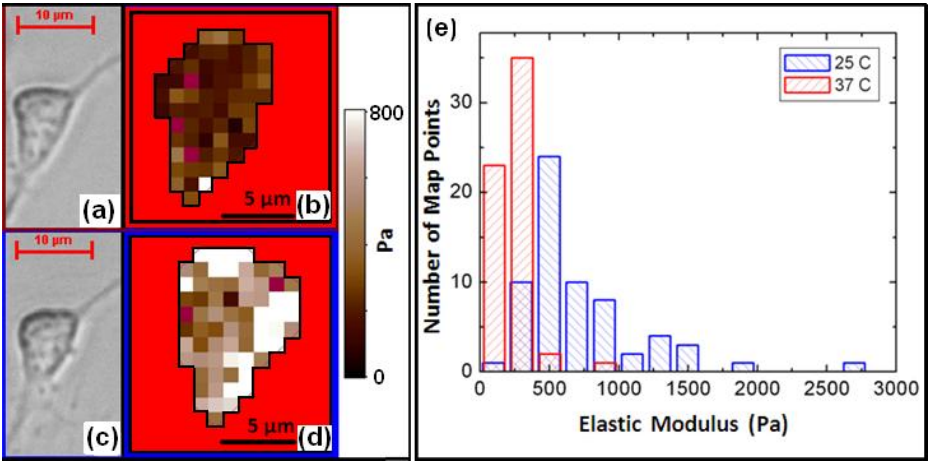

**Figure 2**. Variation of the elasticity maps with temperature. (a) Optical image of a chemically unmodified live cortical cell not undergoing active neurite growth taken at 37ºC. (b) Elasticity map for the cell shown in (a) measured at 37ºC. (c) Optical image of the same cell as in (a) taken at 25ºC. (d) Elasticity map of the same cell as shown in (b) measured at 25ºC. (e) Histograms of elastic moduli for the elasticity maps shown in (b) (data shown in red) and (d) (data shown in blue). The average elastic modulus changes significantly between the two maps, from (239 ± 30) Pa at 37ºC, to (907 ± 80) Pa at 37ºC (the quoted uncertainties represent the standard deviations of the mean). Similar data was obtained for 10 additional unmodified cells.

Next, we monitor the variation in the elastic modulus of the chemically unmodified neuron soma with changes in ambient temperature. Values of elastic modulus from elasticity maps for each cell were averaged (50-100 points per cell, depending on cell size) to give an average soma elastic modulus for the cell at either 37ºC or 25ºC (N=13 cells). Figure 2 shows an example of this type of measurements. Figures 2(a) and 2(b) display the optical images and the elasticity map for a neuron measured at 37ºC, while figures 2(c) and 2(d) show respectively, the optical image and the elasticity map for the same neuron measured at 25ºC. The histograms of measured values for the elastic modulus are given in figure 2(e) (red data points correspond to the map taken at 37ºC, and blue data points correspond to the map taken at 25ºC). The average elastic modulus shows a significant increase (by a factor of 3.8) with a decrease in temperature from 37°C to 25°C.

A summary of the temperature-response obtained for all the cells is shown in figure 3, where we plot the percent-wise change in the average value of the elastic modulus for the cell measured at 25°C with respect to the value for the same cell acquired at 37°C (represented by the 100% black dashes). In the reminder of this section, we focus only on the results obtained for chemically unmodified neurons (red circles in figure 3), while the data for chemically modified cells is discussed in section 3.2). These measurements show that 11 neurons (out of 13) display a significant increase (50-350%) in the average soma elastic modulus when temperature is changed from 37°C to 25°C. We also find that this process is reversible, i.e. all the cells recover the initial values for the elastic modulus upon restoring the temperature to 37°C.

To further demonstrate the reversibility of the process we measure the variation of the elastic modulus through a sweep of temperatures: 25°C, followed by 31°C, followed by 37°C, followed again by 31°C, 25°C, and finally 31°C. An example of this type of measurement is given in figure 4. First, we note that the values of the elastic modulus measured at 31°C fall in between those measured at 25°C and 37°C, respectively. Second, the values of the elastic modulus measured at a given temperature in the cycle are consistent within 10%. Similar results were obtained for 6 additional cells, demonstrating the reversibility of the temperature response.

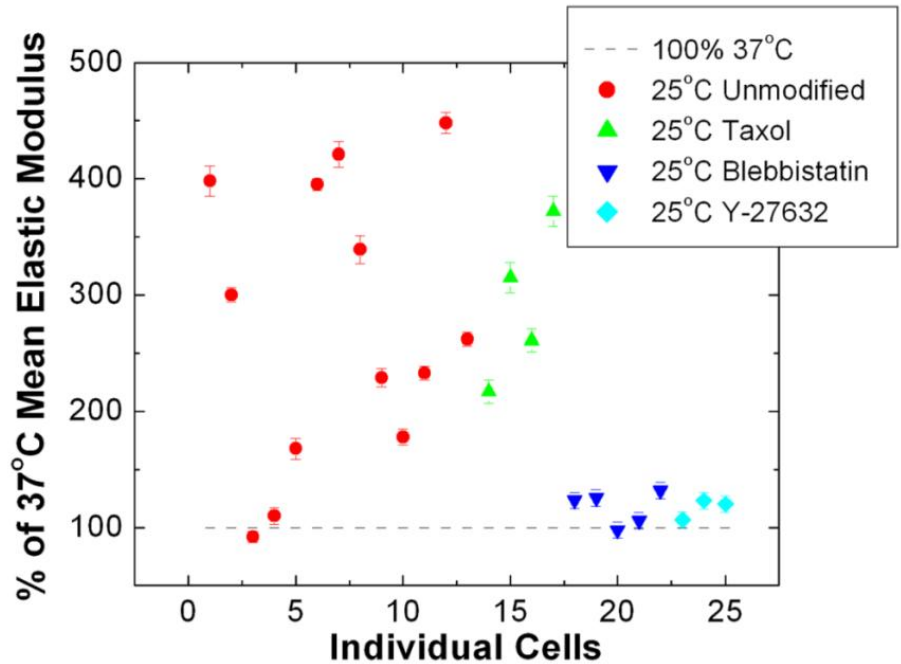

**Figure 3**. Summary of the temperature-induced elasticity changes in neurons. The average elastic modulus of individual cells at 25°C is measured as percent of the average elastic modulus of the same cells at 37°C (represented by the 100% dashed line). The average elastic modulus of each cell was obtained by averaging the cell elasticity map (50-100 points per cell, depending on cell size). Error bars indicate standard error of the mean. Values obtained from chemically unmodified neurons are shown as red circles. Taxol modified cells are shown as green triangles. Blebbistatin modified cells are shown as blue inverted triangles. Y-27632 modified cells are shown as teal diamonds (Taxol, Blebbistatin and Y-27632 data are discussed in section 3.2). Overall, the values for the mean elastic modulus at 37°C range from (160±10) Pa to (595 ± 72) Pa, and the corresponding mean values at 25°C range from (335 ± 21) Pa to (2560±209) Pa. Unmodified cells and Taxol-treated cells at 25°C yield mean values that are statistically different than those measured at 37°C ($p \leq 0.05$ one-way ANOVA for both data sets). Blebbistatin and Y-27632 treated cells yield mean values that are not statistically different than those measured at 37°C ($p \geq 0.5$ one way ANOVA for both data sets), demonstrating that the temperature-induced stiffening effect is suppressed for the cells treated with these two inhibitory drugs.

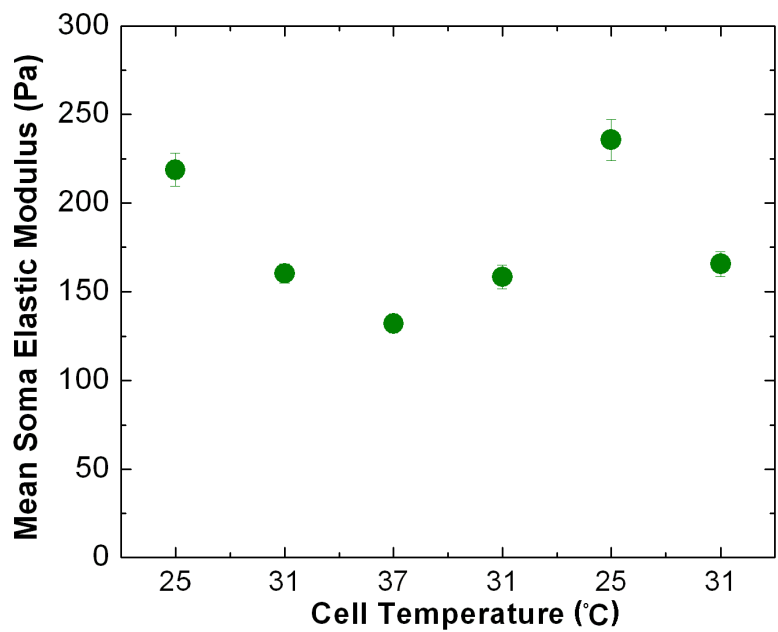

**Figure 4**. Reversibility of the temperature-induced shifts in the elastic modulus. Average soma elastic modulus for an individual cell measured in the sequence of temperatures listed along the x-axis (25°C, 31°C, 37°C, 31°C, 25°C, 31°C). Error bars represent the standard error of the mean. The cell was maintained at each new temperature for a minimum of 10 minutes prior to the measurement. The values of the elastic modulus measured at a given temperature in the cycle are consistent within 10%, demonstrating the reversibility of the temperature response. Similar results were obtained for all cells measured in similar temperature cycles (N=7 cells total).

*3.2 Temperature-dependent response of chemically modified cells*

      In an effort to better understand the observed dependence of the cell elastic modulus on temperature we measured the effect of several drugs that have either a stabilizing or inhibitory effect on the cytoskeleton dynamics. First, we measure the effects of Taxol, a drug that stabilizes microtubules against changes in aggregation and against temperature-induced changes in microtubule flexural rigidity [21, 22]. We obtained elastic modulus maps at 37°C and 25°C of cells treated with 10 μM Taxol (see Supplementary figure S2 for an example). A summary of these results is presented in figure 3: the data represented by green triangles shows that the Taxol-modified cells (N=4) exhibit a significant change in the average elastic modulus with a temperature drop from 37°C to 25°C ($p \leq 0.05$, one-way ANOVA). This variation is within the typical range observed for

unmodified cells (red dots). We conclude that there is no significant difference in the temperature response of Taxol-stabilized cells vs. cells that were not chemically modified.

We have also obtained elastic modulus maps of cells treated with either Blebbistain or Y-27632, two well-known myosin II inhibitors. Blebbistatin inhibits the activity of nonmuscle myosin II motors, and leads to disruption of contractile actin-myosin interactions within the cell [4, 23, 24], and inhibition of myosin II binding to actin in the inactive state [25]. Y-27632 acts on a different pathway than Blebbistatin [25]. Specifically, Y-27632, inhibits phosphorylation of the Myosin Light Chain (MLC), through inhibition of ROCK 1 [28, 29]. Inhibition of MLC phosphorylation decreases myosin II affinity for actin, disrupting actin contraction/ reorganization, and favoring myosin II remaining in the unbound state [30].

Examples of maps of elastic modulus for either Blebbistatin (10 μM) or Y-27632 (10 μM) - treated cells, measured at 25°C and 37°C, as well as a comparison with similar maps obtained on chemically unmodified cells are given in figure 5. Figures 5(a)-(c) show the histogram for the values of the elastic modulus (figure 5(a)), and the elasticity maps at 25°C and 37°C (figures 5(b) and 5(c), respectively) measured for a chemically unmodified cell. Figures 5(d)-(f) show the corresponding data for a Blebbistatin-modified cell, while figures 5(g)-(i) show the corresponding data for a Y-27632 modified neuron. For the chemically unmodified cell (figure 5(a)-(c)) the average elastic modulus increased by a factor of 4 (from (595±72) Pa to (2560 ± 209) Pa), with a decrease in temperature from 37°C to 25°C. In contrast, the Blebbistatin and Y-27632 (figures 5(d)-(i)) data show only a minor increase in the average elastic modulus with a corresponding decrease in temperature from 37°C to 25°C: (399 ± 29) Pa to (423 ± 24) Pa for Blebbistatin, and (210 ± 32) Pa to (223 ± 18) Pa for Y-27632, respectively. Similar results were obtained for N=6 additional cells, treated with either Blebbistatin or Y-27632. A summary of these results is given in figure 3. For the Blebbistatin-modified cells the average elastic modulus varies slightly (between 2-30%) between measurements taken at 37°C and 25°C (figure 3, blue inverted triangles). For the Y-27632-modified cells the data shows that the elastic modulus values differs by less than 25% between measurements at 37°C and 25°C. Taken together, the Blebbistatin and Y-27632 data (N=8 cells) shows that the inhibition of myosin II results in a dramatic decrease in the temperature-induced elastic response observed in chemically unmodified or taxol-treated cells.

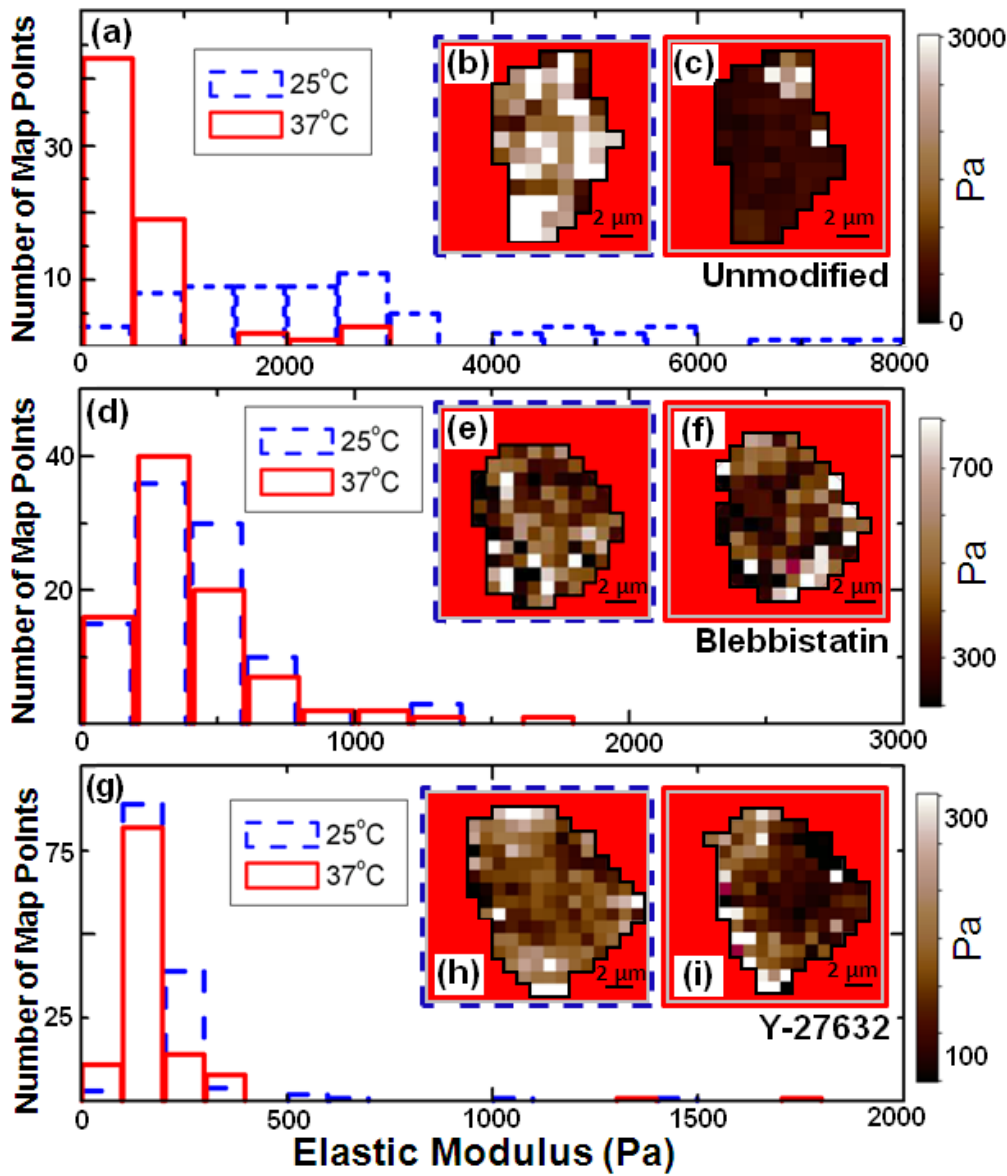

**Figure 5**. Comparison between the temperature-induced elasticity changes measured in unmodified vs. chemically treated neurons. (a) Histograms of the values for elastic modulus for a chemical unmodified neuron measured at 37°C (solid red line) and 25°C (dotted blue line). (b) Elasticity map of a neuron measured at 25°C (histogram shown in (a) as dotted blue line). (c) Elasticity map of the cell shown in (b) measured at 37°C (histogram shown in (a) as continuous red line). (d) Histograms of the values for elastic modulus for a neuron treated with 10 μM of Blebbistatin, and measured at 37°C (solid red line) and 25°C (dotted blue line). The corresponding elasticity maps are shown in (e) (data taken at 25°C), and (f) (data taken at 37°C). (g) Histograms of the values for elastic modulus for a neuron treated with 10 μM of Y-27632, and measured at 37°C (solid red line) and 25°C (dotted blue line). The corresponding elasticity maps are shown in (h) (data taken at 25°C), and (i) (data taken at 37°C). Similar data as in (d)-(i) were obtained for 6 additional cells.

*3.3 Temperature-induced cytoskeletal rearrangements measured by combined AFM and fluorescence microscopy*

We have previously reported that regions of high stiffness in the AFM elastic modulus maps of healthy cortical neurons measured at 37°C correlate to regions of increased concentration of the cytoskeletal component tubulin as observed through fluorescence imaging [6]. For example, figure 6(a) shows a fluorescence image of a live cortical neuron stained for tubulin, while figure 6(b) shows the elasticity map of the same neuron measured at 37°C. Figure 6(a) shows that the region of high microtubule concentration (i.e. high fluorescence intensity) along the top of the cell overlaps with the region of high elastic modulus displayed in figure 6(b). We have consistently observed this effect for all the tubulin-stained cells (N=12) measured at 37°C, with an overall 70-100% overlap between the areas of high elastic modulus (i.e. values of elastic modulus above the average value for the cell) and the regions of high tubulin density (i.e. fluorescence intensity higher than the average value for the cell).

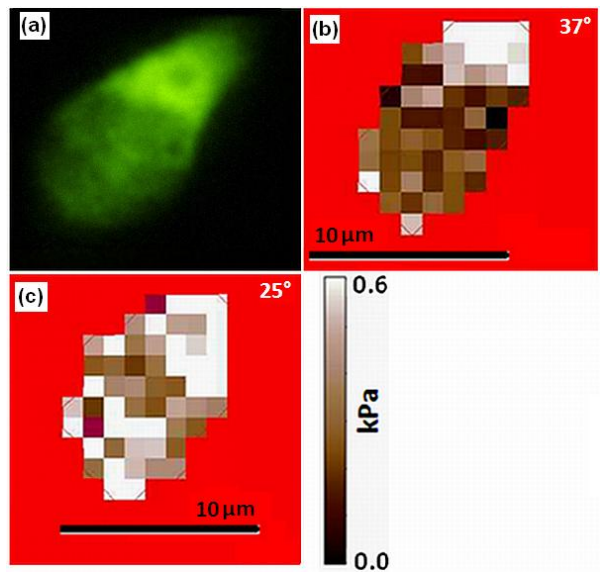

**Figure 6**. Correlation between tubulin density and cell stiffness at different temperatures. (a) Fluorescence image of a cortical neuron soma stained for tubulin. No change in the fluorescence image was observed between 37°C and 25°C. (b) Elasticity map of the cell shown in (a) taken at 37°C before the cell was stained for tubulin. The data shows a large overlap between tubulin-dense regions and high stiffness areas of the cell at 37°C. (c)

Elasticity map of cell shown in (a) taken at 25°C. The overlap between high-tubulin concentration and high stiffness areas is greatly reduced at 25°C. Figures (b) and (c) have the same scale bar for the elastic modulus.

Here, we find that the correlation between regions of high-tubulin concentration and the high-stiffness areas of the soma decreases dramatically as the temperature is dropped from 37°C to 25°C. For example, figure 6(c) displays the elasticity map measured at 25°C for the same neuron as in figures 6(a) and 6(b). We note that the fluorescence image did not show any measurable change in the location of regions of high and low tubulin concentration inside the cell soma, upon the temperature drop from 37°C to 25°C. However, a comparison between figure 6(a) and figure 6(c) shows that at 25°C, there are many regions of the cell displaying high stiffness (center and bottom of the cell) that no longer overlap with regions of high tubulin concentration. From figure 6 and similar data taken on N=12 cells, we measure a significant decrease in the overall overlap between the regions of high tubulin concentration and the areas of high elastic modulus taken at 25°C. A detailed analysis performed on all the measured cells shows that the overlap between these two regions is between 30-50% at 25°C, compared to 70-100% overlap measured at 37°C. We conclude that overall, the highest stiffness regions no longer correlate directly with regions of high microtubule concentration at 25°C.

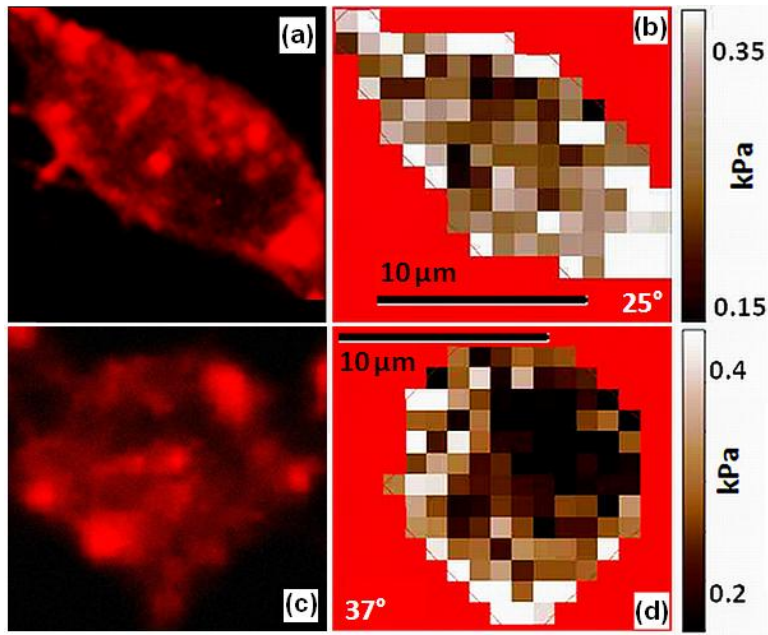

**Figure 7**. Correlation between actin density and cell stiffness at different temperatures. (a) Fluorescence image of a cortical neuron soma stained for actin and taken at 25°C. (b) Elasticity map of cell shown in (a) measured at 25°C before the cell was stained for actin. The data shows a large overlap between actin-dense regions and high stiffness areas of the cell at 25°C. (c) Fluorescence image of a different cortical neuron soma stained for actin and taken at 37°C. (d) Elasticity map of cell shown in (c) measured at 37°C before the cell was stained for actin. The overlap between high-actin concentration and high stiffness areas is greatly reduced at 37°C.

To explore the observed change in elasticity with temperature, we have also stained neuronal cells for actin concentration (figures 7(a) and 7(c)) and map the cell elastic modulus at 25°C (figure 7(b)) and 37°C (figure 7(d)). Figure 7 shows that at 25°C there is about 75% overlap between the regions of high actin concentration (i.e. high fluorescence intensity in figure 7(a)) and the regions of high elastic modulus (displayed in figure 7(b)). However, this correlation is no longer observed for cells measured at 37°C, where there is less than 30% overlap between high-actin concentration and high elastic modulus areas (figures 7(c) and 7(d)). Similar results were found for N=6 additional cells.

The combined fluorescence/AFM measurements strongly suggest that there is a shift in the primary cytoskeletal component responsible for high elastic modulus of the neuron soma, from tubulin to actin dominated regions, with a drop in temperature from 37°C to 25°C. This hypothesis is confirmed by experiments performed on the *same* cell, stained for both microtubules and actin (figure 8). The combined elasticity/fluorescence maps show a significant overlap between cellular regions with high *tubulin* concentration and regions displaying higher than average elastic modulus, when measurements are taken at 37°C (figures 8(a) and 8(b)). Furthermore, measurements of the same cell at 25°C show a large overlap between areas of high elastic modulus and regions of high *actin* concentration (figures 8(c) and 8(d)). No change in the fluorescence image for tubulin (figure 8(a)) was observed when temperature was changed between 37°C and 25°C. Overall, these results demonstrate a temperature-induced shift in the soma elasticity from tubulin-dominated stiffness at 37°C to actin-dominated stiffness at 25°C.

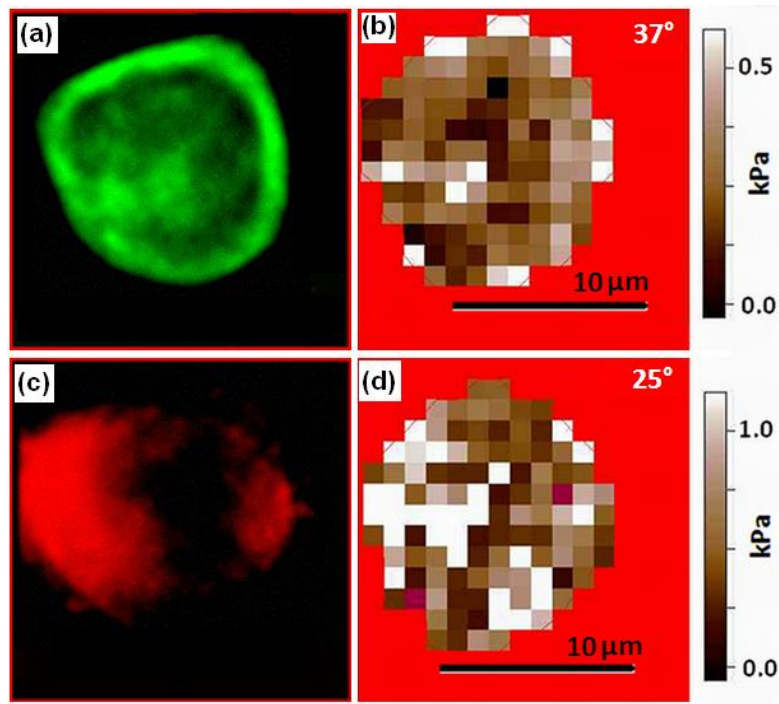

**Figure 8**. Cytoskeletal components responsible for high elastic modulus of the neuron soma at different temperatures. (a) Fluorescence image of a cortical neuron soma stained for tubulin (no change in the fluorescence image for tubulin was observed between 37°C and 25°C). (b) Elasticity map of cell shown in (a) measured at 37°C. Higher than average values of elastic modulus at 37°C correlate with regions of high-actin concentration. (c) Fluorescence image of the same cortical neuron soma as in (a) stained for actin and taken at 25°C. (d) Elasticity map of cell shown in (c) measured at 25°C. There is a large overlap between high-actin concentration and high stiffness areas at 25°C.

We have additionally examined the data from the chemically unmodified cells at 25°C and 37°C to determine if there is any preference for temperature-dependent stiffening between the outer and the inner regions of the soma. First, we isolate the points on the elasticity map confined to the outer 20% of the soma radius and calculate the average value of elastic modulus for these points. Next, we calculate the corresponding average values for the elastic modulus of the remaining points of the elasticity maps (from the center of the cell up to 80% of the soma radius). We then compare the percent shift in these two average values of the elastic modulus as the cell temperature was decreased from 37°C to 25°C. The results (see figure S3 in the supplementary materials) show similar contributions from both regions, with no preferential shift in the elasticity of the soma periphery vs.

the central regions. This indicates that the observed temperature-induced stiffening is a global effect, with the whole soma undergoing the observed variation in elasticity with temperature.

## 4. Discussion

In this study we combined AFM elasticity mapping with fluorescence microscopy to measure the effects of temperature on the elastic properties of cortical neuron soma. Our results show a significant increase in the cell elastic modulus with a decrease in temperature between 37°C to 25°C (figures 2, 3, 4 and 5(a)). We also find that that this effect is reversible, i.e. the elastic modulus recovers the initial value with a subsequent increase in temperature from 25°C to 37°C.

Furthermore, our data shows that while tubulin dense regions dominate the elastic landscape at 37°C, they do not play a significant role in the observed increase in cell elastic modulus at 25°C. This result is supported by both fluorescent/AFM (figure 6) and Taxol experiments (figure 3 and supplementary figure S2). We note that if changes in microtubule aggregation, stability, and rigidity were primarily responsible for the observed shifts in elasticity with temperature, we would expect that the effect of temperature on the cell stiffness should be significantly reduced in cells cultured with Taxol (which reduces microtubule dynamics [6, 21, 22]). However, our measurements do not show a substantial difference in the temperature response of Taxol-stabilized cells vs. cells which were not chemically modified (figure 3). The data shows instead a significant *increase* in stiffness of the Taxol treated cells ($p \leq 0.05$, one-way ANOVA) with a temperature drop from 37°C to 25°C, which is similar to the effect measured on chemically unmodified cells. Taken together, the results obtained for Taxol treated neurons and the measurements that show no significant correlation between the tubulin dense regions and the high elasticity areas of the cell at 25°C indicate that the observed stiffening effect was not produced by changes in the microtubule component of the cytoskeleton.

This conclusion is further supported by the results obtained on the cells stained for actin. While neurons measured at 37°C display only a small overlap between areas of high elastic modulus and regions of high actin concentration (figures 7(c) and 7(d)), we find that the overlap between these two regions is greatly increased at 25°C (figures 7(a) and 7(b), and figures 8(c) and 8(d)).

The neuronal cytoskeleton is a complex biopolymer network, which assembles a multitude of actin filaments, intermediate filaments and microtubules in order to guide intracellular transport, regulate the cell shape and mechanical stability, and control the extension of neurites. Assemblies based on F-actin filaments are typically the dominant mechanical structures within most living cells. Some of these structures form semi-flexible contractile networks where F-actin filaments are cross-linked by a variety of binding proteins (such as α-actinin linkers, β-spectrin, filamins, myosin II motors etc.). These actin binding proteins control the length of the F-actin filaments, crosslink filaments with each other, apply forces and regulate the mechanical stresses along the filaments. In particular, contractile stresses arising from interactions between myosin II motors and actin filaments play an important role in the elastic properties of the cytoskeleton.

Myosin II motors generate both compressive and tensile stresses on F-actin, which could result in sliding of the F-actin filaments past one another [31-34], variations in the bending rigidity of the F-actin filaments [31, 32], and changes in the persistence length and flexibility of F-actin cytoskeletal networks [33, 35-37]. For example, *in vitro* studies on purified actin networks cross-linked by filamin A have shown that adding myosin II motors leads to a dramatic increase in the network elastic modulus [34]. These results demonstrate that one function of myosin II is to pull the actin filaments and produce internal stresses that stiffen the cytoskeleton. However, the effects of myosin II on actin filaments are dependent on many factors including the presence, concentration, and type of cross-linkers [34] [38], metabolic activity of the myosin II motors (dependent on the temperature and availability of ATP[39-41]), as well as length and orientation of actin filaments [39, 42]. For example, for isolated actin/myosin II mixtures an increase in myosin II activity has been shown to fluidize the network by producing a compliant fluid-like state, and thus to lower the network elastic modulus [39, 40]. In these experiments, inhibition of myosin II activity (through substitution of adenosine triphosphate for adenosine diphosphate in solution) has resulted in an increase in the density of crosslink F-actin filaments due to the binding of inactive myosin II motor domains, and thus resulted in a more solid-like consistency of the actin-myosin system [39]. These results provide a possible scenario for our results which show a decrease in neuron stiffness with increasing temperature: since myosin II activity increases with increasing temperature [43, 44], and active myosin II fluidizes the cells while inactive myosin II acts as a cross linker for F-actin, we expect that an increase in activity of the myosin II at 37°C results in a decrease in the average elastic modulus of the cell. This scenario is also consistent with the Blebbistatin and Y-27632 results, as these drugs block the myosin II in the active (detached) state[45], and therefore it reduces its cross linking activity, resulting in a suppressed variation of cell stiffness with temperature.

Our results are also consistent with several *in vitro* experiments, performed on simpler cross-linker/F-actin networks in the absence of myosin II motors [46-49]. These studies have used variations in temperature to systematically change the binding affinity of actin binding proteins (in these studies α-actinin) to F-actin, and thus to measure the effect of temperature on the mechanical properties of these networks [46-49]. These experimental results show that as temperature is *increased* the F-actin networks cross-linked with α-actinin become *softer* and more fluid-like, i.e. the main effect associated with increasing temperature in these systems is an increase in the unbinding rate of the actin binding proteins from F-actin. This results in a more dynamic network that can relax stress as the cross-link dissociation rate increases. For example, the increase in elastic modulus for an isolated F-actin and α-actinin gel has been shown to be on the order of 50% with a drop in temperature from $37°C$ to $25°C$ [46].

These findings are in agreement with our results, which show that the elastic modulus of the neuron soma exhibits only a slight increase when myosin II is inhibited through the action of Blebbistatin or Y-27632 (figures 3 and 5). Blebbistatin has been shown to disrupt both the cross-linking activity of myosin II and the contraction of F-actin structures through inhibition of myosin II [4, 23, 24, 45]. Similarly, Y-27632 inhibits the Rho kinase pathway leading to a dampening of myosin II dynamics in the soma [13, 25-27]. Our data shows that by chemically suppressing myosin II activity/cross-linking in cortical neurons through Blebbistatin or Y-27632, individual neurons undergo only a slight increase in stiffness when the temperature is decreased from $37°C$ to $25°C$, with an average change in elastic modulus over the whole soma of less than 30%. This value is consistent with a decrease in F-actin/α-actinin cross linking with increasing temperature, as discussed above. We also note that the temperature-induced changes in entropic elasticity of flexible polymer networks predict a linear *increase* of the elastic modulus with temperature [50]. However, in the temperature range used for our studies the changes in elastic modulus due to entropic elasticity are negligible (less than 4%).

Additionally, our data shows that when suppression of myosin II was not controlled through the application of Blebbistatin or Y-27632, the cell elastic modulus displayed a significant increase (with the average stiffness of the soma increasing between 50-350%) with temperature decreasing from 37°C to 25°C. We also find that the high stiffness values at 25°C correspond to the actin-dense regions of the cell soma. These results are also in agreement with recent experiments [31-33, 36] which demonstrate that myosin II motors, acting as F-actin cross linkers, can cause large bending fluctuations in F-actin networks, resulting in an effective reduction of the chain length $L_e$, and in the effective persistence length $l_p$ of the network. The approach most commonly used to model the mechanical properties of the cells is to consider the cytoskeleton as a soft, semi-flexible viscoelastic network, with the elastic modulus dominated by the bending rigidity, the length and the crosslinking properties of the biopolymer filaments that form this network [33, 34, 37, 51]. Within this general model it can be shown that the elastic modulus $G$ scales as [51]:

$$G \sim \frac{1}{k_B T}\left(\frac{\kappa^2}{\xi^2 L_e^3}\right) \qquad (1)$$

where, $k_B$ is the Boltzmann constant, $T$ is the temperature, $\kappa$ is the chain bending modulus, and $\xi$, and $L_e$ represent the characteristic mesh size of the network and the typical length of a chain segment, respectively. Both $\xi$ and $L_e$ decrease with increasing concentrations of chains, and for a densely cross-linked network we have that [51]:

$$L_e \approx \xi < l_p \qquad (2)$$

where $l_p$ is the persistence length of the chain.

Thus, from equations (1) and (2) above we expect an increase in the elastic modulus of an F-actin network, with an increase in bending and in the density of crosslink chains, resulting from a decrease in the activity of actin-detached myosin II with decreasing temperature. Our observed increase in the elastic modulus suggests a myosin II-induced power law dependence of the F-actin network mesh size $\xi$, and length chain $L_e$ with temperature $T$. From the data presented in figure 4 (and similar data obtained on N=6 cells), and equations (1) and (2), we obtain:

$$L_e \approx \xi \sim T^{(3.73 \pm 2.1)} \qquad (3)$$

This result is consistent with both rheology and AFM measurements on various cell types [13, 52], which have shown that the cell elastic modulus exhibits a power-law behavior over a relatively large temperature range (13°C to 37°C). We also note that the paper by Sunyer and al. [13] has found a power law behavior but an opposite trend, i.e. an *increase* in the elastic modulus with increasing temperature. This suggests that for other types of cells (and at least in the case of the more flexible/elastically compliant alveolar epithelial cells) variations in temperature could have a different effect on the cell elastic properties than in cortical neurons. Indeed, as it has been shown by several *in vitro* experiments [39, 40] the interplay between myosin II induced filament sliding and cross linking is rather complex, depending on several factors such as length and orientation of F-actin filaments, stress relaxation rate, type, concentration and activity of other types of cross linking proteins etc. Future temperature-dependent experiments performed on other types of neurons (such as peripheral Dorsal Root Ganglia), as well as other cell types could further elucidate the full mechanisms that regulate the interplay between passive crosslinking and molecular motor activity, as well as their role in cytoskeletal dynamics.

## 5. Conclusions

We have performed combined AFM-elasticity, fluorescence microscopy and temperature-dependent experiments on life cortical neurons. Our results demonstrate a significant increase in the average soma elastic modulus with a decrease in temperature in the range 37°C to 25°C, and we identify the cytoskeletal components associated with these changes in cellular elasticity. We also show that this process is reversible with the cell elastic modulus returning to the initial values upon a return of the cell to 37°C, and that there is a shift in the elastic energy landscape of the neuron soma, from tubulin dominated regions at 37°C to actin dominated regions at 25°C. We demonstrate that the observed increase in soma stiffness with decreasing temperature is significantly reduced if the activity of myosin II motors is suppressed via cellular modification with either Blebbistatin or Y-27632.

These results present a potential method to reversibly control the elastic properties and the actomyosin activity in cortical neurons, through variations in temperature without requiring additional chemical modifications. *In vitro*, such temperature-dependent effects may be useful as controllers of nerve functions to allow further study of cell physiology, biomechanics and motility. Furthermore, combined temperature-dependent elasticity mapping with chemical modifications experiments could lead to a detailed understanding of actin/myosin II dynamics in neurons *in vivo*, thus offering insight into nerve physiology under normal vs. disease or stress conditions, or into the fundamental cell functions and repair mechanisms. These temperature-related effects are also important during tissue damage or infections. Finally, our results demonstrate that the biomechanical properties of cells measured at 37°C (close to physiological conditions for many mammals and birds) are significantly different than the corresponding properties measured at 25°C (which is the case for many results reported in literature). A comprehensive study of various cell types and a larger range of temperatures (including higher temperatures that simulate disease/fever conditions) may yield further insight into the particular properties of the actin cytoskeleton which define the direction of a cells temperature response, as well as the effect of temperature-induced cellular damage on the mechanical properties of neurons. It may also be of interest to examine the temperature response of neuronal cells from a mature animal, as the cytoskeletal response of neurons in the developing (embryonic) brain may be different than those in the fully developed state. Future work may also include studies into how changes in temperature impact cellular adhesion and substrate-elasticity-preferences, which might be affected by actin and myosin dynamics.

**Acknowledgments**

The authors thank Dr. James D. White and Dr. Min Tang-Schomer for their assistance and technical guidance on the neuron culture methods. We also thank Dr. Steve Moss's laboratory at Tufts Center of Neuroscience for providing embryonic rat brain tissues. This work was supported by the National Science Foundation (grant CBET 1067093). We would also like to thank one of the reviewers for providing insightful suggestions about the possible scenarios for the observed temperature dependence.